\documentclass[fleqn,10pt]{wlscirep}

\title{Finite temperature physics of $1D$ topological Kondo insulator: Stable Haldane phase, Emergent energy scale and Beyond}

\author[1,*]{Yin Zhong}
\author[2,3]{Yu Liu}
\author[1]{Qin Wang}
\author[4]{Ke Liu}
\author[2,3]{Hai-Feng Song}
\author[1,5]{Hong-Gang Luo}
\affil[1]{Center of Interdisciplinary Studies $\&$ Key Laboratory for Magnetism and Magnetic Materials of the Ministry of Education, Lanzhou University, Lanzhou 730000, China}
\affil[2]{Institute of Applied Physics and Computational Mathematics, Beijing 100088, China}
\affil[3]{Software Center for High Performance Numerical Simulation,China Academy of Engineering Physics, Beijing 100088, China}
\affil[4]{Arnold Sommerfeld Center for Theoretical Physics, University of Munich, Theresienstrasse 37, 80333 Munich, Germany}
\affil[5]{Beijing Computational Science Research Center, Beijing 100193, China}
\affil[*]{zhongy@lzu.edu.cn}

\begin{abstract}
In recent years, interacting topological insulators have emerged as new frontiers in condensed matter physics, and the hotly studied topological Kondo insulator is one of such prototypes. Although its zero-temperature ground-state have been widely investigated, the finite temperature physics on topological Kondo insulator is rarely explored. Here, as an example, we study the finite temperature properties of one-dimensional $p$-wave periodic Anderson model with numerically exact determinant quantum Monte Carlo simulation. It is found that the topological Haldane phase established for ground-state is still stable against small thermal fluctuation and its characteristic edge magnetization develops at low temperature. Moreover, we use the saturated low-$T$ spin structure factor and the $\frac{1}{T}$-law of susceptibility to detect the free edge spin moment, which may be relevant for experimental explorations. We have also identified an emergent energy scale $T_{cr}$, which signals a crossover into interesting low-$T$ regime and seems to be the expected Ruderman-Kittel-Kasuya-Yosida coupling. Finally, the collective Kondo screening effect has been examined and it is heavily reduced at boundary, which may give a fruitful playground for novel physics beyond the well-established Haldane phase and topological band insulators.
\end{abstract}
\begin{document}

\flushbottom
\maketitle
%
%
\thispagestyle{empty}

\section*{Introduction}
In recent years, topological states of matter has become one of most important issues in condensed matter physics, motivated by intensive and successful investigations on time-reversal invariant topological insulator.\cite{Hasan2010,Qi2011,Hohenadler2013,Bansil2016,Witten2016,Chiu2016,Senthil2015,Liu2016,Dzero2016}
To date, most of real-life topological insulator-like materials can be well described in terms of topological band theory,\cite{Bansil2016}
i.e. their basic electronic structures can be readily understood with the framework of non-interacting single-electron picture.

However, the well-known topological Kondo
insulator candidate, Samarium Hexaboride (SmB$_{6}$) is an exception,\cite{Dzero2016,Dzero2010} which opens up a new window to correlated topological insulators.\cite{Hohenadler2013}
For example, the large Fermi velocity observed in its surface state suggests a radical breakdown of collective Kondo entanglement/screening on surface state,\cite{Jiang2013,Neupane2013,Alexandrov2015} which embodies the concept of surface Kondo breakdown. Furthermore, high-field quantum oscillation measurement triggers a possibility of a neutral three-dimensional Fermi surface in such bulk insulator,\cite{Tan2015,Baskaran2015} and inspiring theoretical explanations with new fractionalized quasi-particle, e.g. free Majorana fermion in Majorana Fermi sea and failed superconductor (Skyrme insulator),\cite{Baskaran2015,Erten2017} spin-one-half spinon in fractionalized Fermi liquid and composite exciton in Fermi liquid.\cite{Thomson2016,Chowdhury2017,Sodemann2017} We note that due to intrinsic strong electron correlations, these plausible proposals/explanations are far from being understood, thus insights from different aspects like exact numerical simulations are highly desirable.

Very recently, we have taken a step on this important issue with a numerically exact zero-temperature quantum Monte Carlo study. We focus on the $1D$ $p$-wave periodic Anderson model,\cite{Zhong2017} which serves a simplified model to understand electron correlation effect on topological Kondo insulator in one spatial dimension.
In this model, the celebrated Haldane phase has been found and the surface Kondo breakdown mechanism relevant to explanation of exotic features in SmB$_{6}$ is critically examined.\cite{Zhong2017,Lisandrini2017} Although realization of zero-temperature surface Kondo breakdown is found to be difficult, it has been argued that the elevated temperature has the potential to destroy Kondo screening.\cite{Alexandrov2015}
Therefore, a finite temperature study on corresponding model has the potential to inspect this novel idea.

At the same time, it is also interesting to investigate the finite temperature effect on the stability of topological Haldane phase, which is rarely done in literature but to face any realistic topological materials, it is essential to consider the thermal fluctuation effect. Also, one may wonder what is the difference between such Haldane-like topological Kondo insulator and the conventional Kondo insulator at finite temperature.

In this work, we study above interesting finite temperature physics on the $1D$ $p$-wave periodic Anderson model. Specifically, we focus on the stability of Haldane phase under thermal fluctuation and explore the fate of its characteristic edge magnetization when elevating temperature. To detect the edge magnetization, the spin structure factor and static spin susceptibility are examined, which is useful for experimental confirmations. Surprisingly, when we compare the $p$-wave model with more conventional $s$-wave hybridized periodic Anderson model, an emergent crossover temperature $T_{cr}$ can be identified. This may relate to the intrinsic Ruderman-Kittel-Kasuya-Yosida (RKKY) interaction for heavy fermion system and agrees with phenomenological heavy fermion two-fluid theory.\cite{Yang2016} In addition, we explore the possibility of the mentioned temperature-driven surface Kondo breakdown. We believe our work is useful for understanding novel topological phases in heavy fermion compounds like SmB$_{6}$ and other interacting topological states of matter.

\section*{Model Hamiltonian and Numerical Method}\label{sec_2}
\begin{figure}[!tb]
  \centering
  \includegraphics[width=0.5\columnwidth]{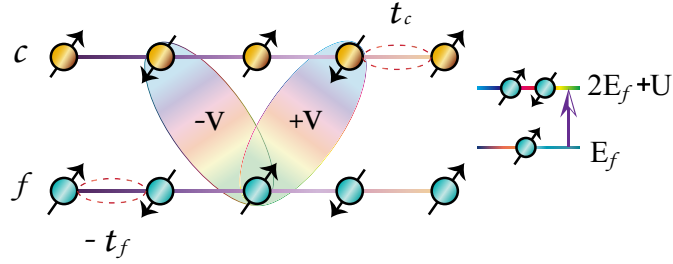}
\caption{$1D$ $p$-wave periodic Anderson model describes a $p$-wave-like hybridization $\pm V$ between local electron orbital (blue) and its neighboring conducting charge carrier (yellow).
The singly occupied local electron has energy $E_{f}$ while double occupation has extra Coloumb energy $U$. The conduction (local) electron hops $t_{c}$ ($-t_{f}$) between nearest-neighbor sites.}
\label{fig:PAM}
\end{figure}

The $1D$ $p$-wave periodic Anderson model we will study is defined as follows:\cite{Zhong2017,Coleman2015}(See Figure~\ref{fig:PAM})

\begin{eqnarray}
&&H_{\mathrm{p-wave}}=H_{c}+H_{f}+H_{p}\nonumber\\
&&H_{c}=\sum_{j\sigma}[t_{c}c_{j\sigma}^{\dag}c_{j+1,\sigma}-t_{f}f_{j\sigma}^{\dag}f_{j+1,\sigma}+\mathrm{H.c.}]\nonumber\\
&&H_{f}=E_{f}\sum_{j\sigma}f_{j\sigma}^{\dag}f_{j\sigma}+U\sum_{j}f_{j\uparrow}^{\dag}f_{j\uparrow}f_{j\downarrow}^{\dag}f_{j\downarrow}\nonumber\\
&&H_{p}=\frac{V}{2}\sum_{j\sigma}[(c_{j+1,\sigma}^{\dag}-c_{j-1,\sigma}^{\dag})f_{j\sigma}+{\rm H.c.}]\label{eq1}.
\end{eqnarray}

Here, $t_{c}$ and $t_{f}$ are the nearest-neighbor-hopping strengths for conduction electron and local $f$-electron. $E_{f}$ denotes the energy level of local $f$-electron, which also has the conventional Hubbard on-site interaction ($U$-term).
The $p$-wave hybridization between conduction and local electron is encoded by the term $H_{p}$, in which the coupling of conduction and local electron is non-local such as the non-trivial spin-orbit coupling between $d$-like conduction electron and $f$-like local electron. Such non-local hybridization gives rise to $1$D $Z_{2}$ topological insulator or Haldane-like phase when Hubbard interaction $U$ is turning on.\cite{Zhong2017,Lisandrini2017} Conversely, the standard $s$-wave periodic Anderson model reads,

\begin{eqnarray}
&&H_{\mathrm{s-wave}}=H_{c}+H_{f}+H_{s},~~~~H_{s}=V\sum_{j\sigma}(c_{j\sigma}^{\dag}f_{j\sigma}+f_{j\sigma}^{\dag}c_{j\sigma})\label{eq2}.
\end{eqnarray}
where the more familiar $s$-wave on-site hybridization $H_{s}$ is assumed.
Formally, these two models have different $c$-$f$ hybridization styles and we will see the resulting physics in the following sections.

It should be emphasized that the zero-temperature ground-state physics of the half-filled $p$-wave model $H_{\mathrm{p-wave}}$ has been studied in our previous work,\cite{Zhong2017} followed by an independent density-matrix-renormalization-group calculation.\cite{Lisandrini2017} The main finding is that the Haldane phase is indeed the ground-state when the on-site Coulomb interaction $U$ is turned on. The smoking gun feature of such state is the decoupled $S=1/2$ edge magnetic local moment, which is observed in the site-resolved magnetization.

To study the finite temperature physics of $H_{\mathrm{p-wave}}$, the method of choice is the state-of-art determinant quantum Monte Carlo (DQMC) simulation. It is noted that DQMC has been widely used in the simulation of correlated electron problems,\cite{Blankenbecler1981,Hirsch1985,Santo2003} including the conventional periodic Anderson model.\cite{Vekic1994,Jiang2014,Lin2015,Hu2017}(Here, $H_{\mathrm{s-wave}}$ is such an example.) However, DQMC simulation on lattice fermion models generally suffer from fermion-minus-sign problem, where effective Boltzmann weight is negative or even complex-valued, which invalids the Monte Carlo importance sampling.\cite{Santo2003} Fortunately, the symmetric half-filling Hubbard and periodic Anderson model are free of minus-sign problem, thus allows a reliable simulation at low temperature and strong coupling.\cite{Gubernatis2016}

For our model $H_{\mathrm{p-wave}}$, the requirement of non-sign problem is $E_{f}=-U/2$ and the chemical for both conduction and f-electron should be set to zero. This leads to a symmetric half-filled system, and the average electron number per site is fixed to $n_{c}=n_{f}=1$.
Furthermore, to compare with previously studied zero-temperature results, we choose open boundary condition and other parameters are set to $t_{c}=t_{f}=V=1$ and $U=2$ with total lattice site number $L=20$. (Much longer chain has been tested as well and no noticeable changes have been found.) Additionally, we use $40000$ sweeps for warmup and $420000$ sweeps for
measurements, the imaginary time interval is $\Delta\tau=0.1$ and statistical errors in our DQMC simulation are typically smaller than the symbol size in the plot and will not be explicitly shown in figures.

\subsection*{Benchmark: compare with $T=0$ case}
To justify the validity of our DQMC simulation, it is crucial to compare this finite temperature method with the zero temperature projector quantum Monte Carlo (PQMC) used in our previous work.\cite{Zhong2017}

\begin{figure}[!tb]
  \centering
  \includegraphics[width=0.3\columnwidth]{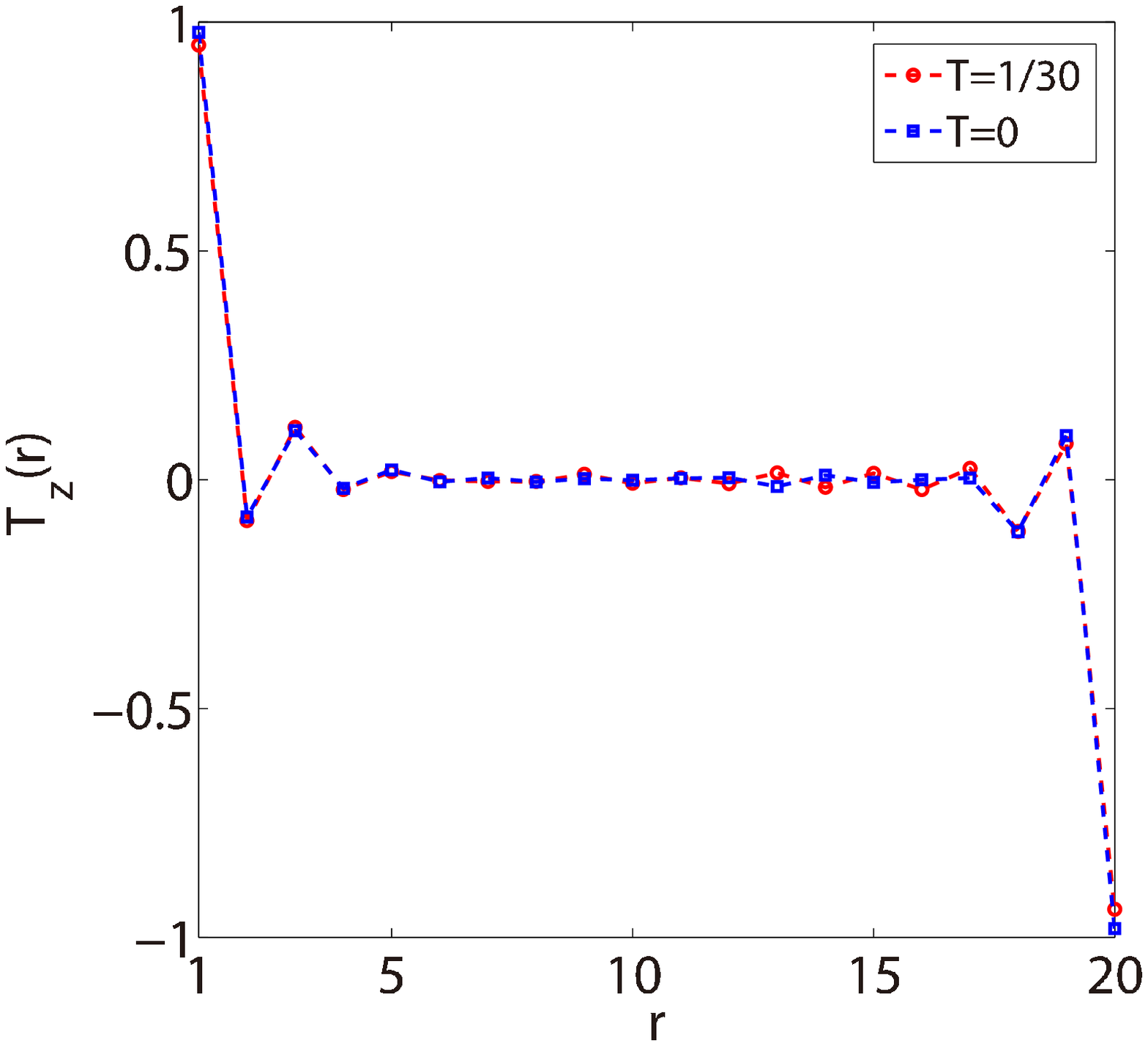}
  \includegraphics[width=0.3\columnwidth]{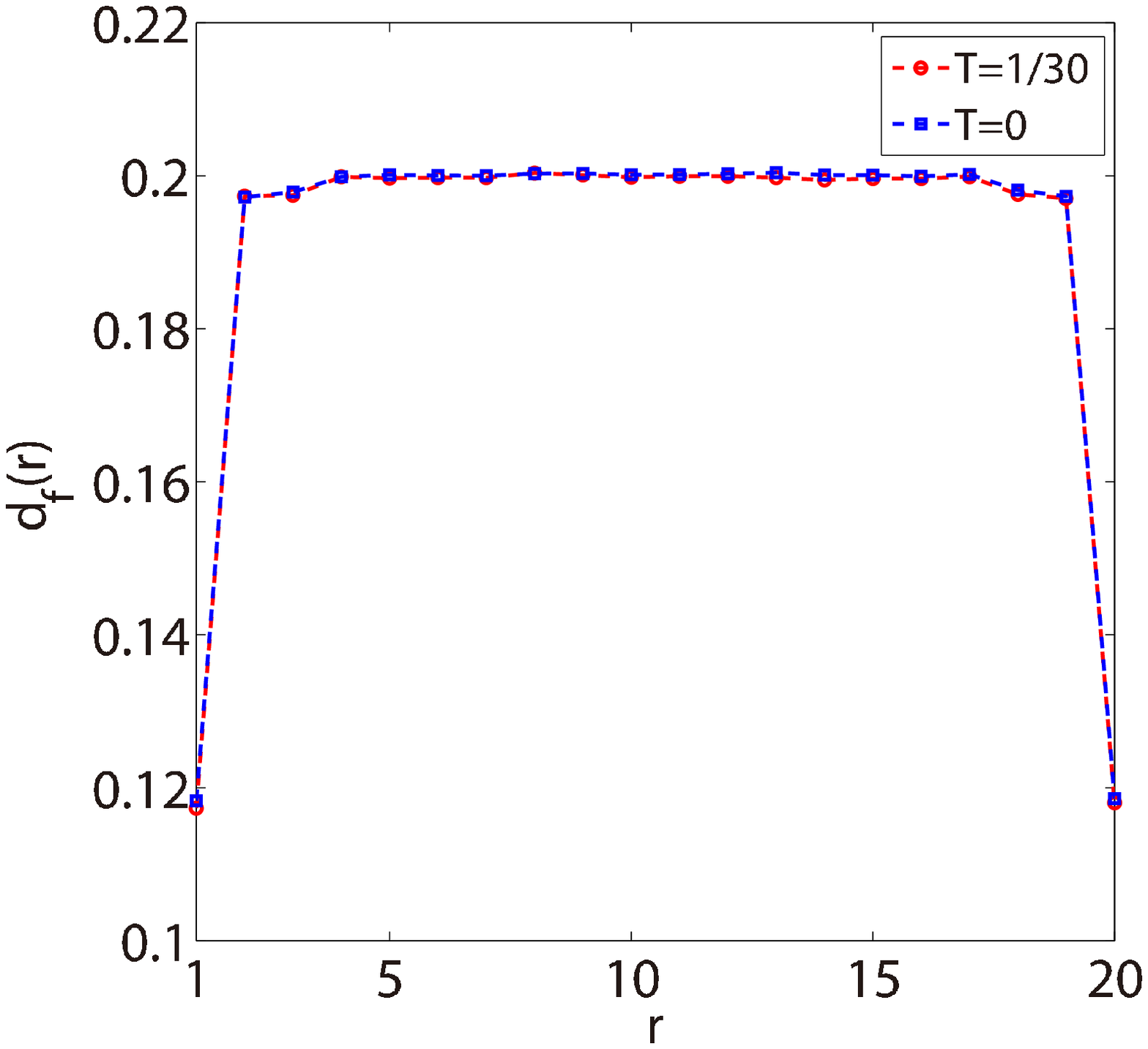}
  \includegraphics[width=0.3\columnwidth]{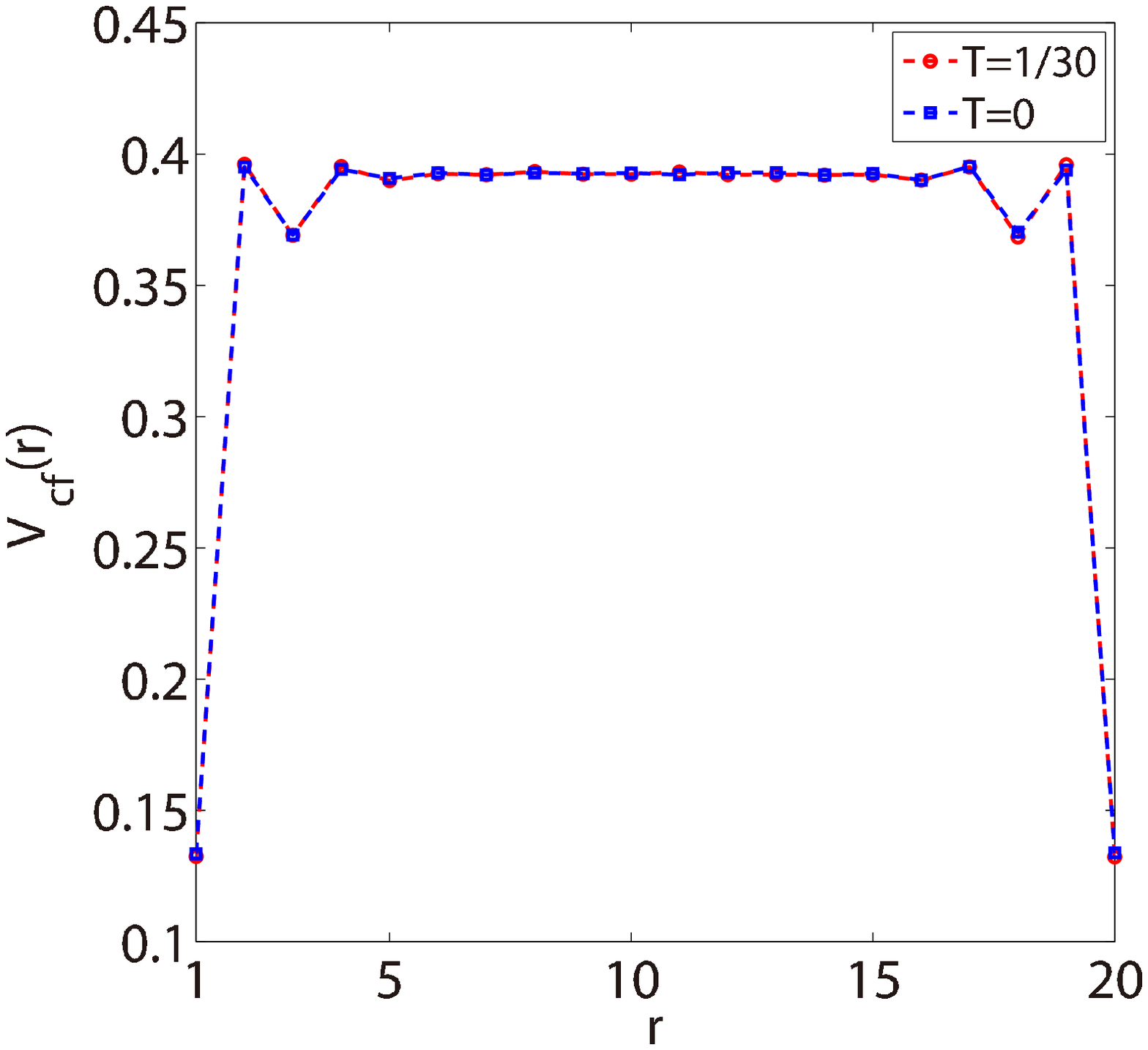}
\caption{Comparison between finite (red) temperature $(T=1/30)$ physical quantities by DQMC and zero (blue) temperature ones by PQMC for $p$-wave periodic Anderson model.\cite{Zhong2017} (Left) Site-resolved magnetization $T_{z}(r)$; (middle) site-resolved double $f$-occupation number $d_{f}(r)$; (right) site-resolved $c$-$f$ hybridization $V_{cf}(r)$. The agreement between DQMC and PQMC is fairly good and this justifies the DQMC simulation algorithm for finite-$T$.}
\label{fig:compare}
\end{figure}

In Figure~\ref{fig:compare}, we have shown typical data calculated from quantum Monte Carlo simulation at finite temperature $T=1/30$ and zero temperature case, for several physical quantities. To be specific, they are site-resolved magnetization $T_{z}$, double occupation number of $f$-electron $d_{f}$ and $c$-$f$ hybridization $V_{cf}$. Here, $T_{z}$ is defined as the total $z$-component magnetization on each site $T_{z}(r)=2\langle S_{z}^{f}(r)\rangle+2\langle S_{z}^{c}(r)\rangle=\langle f_{r\uparrow}^{\dag}f_{r\uparrow}-f_{r\downarrow}^{\dag}f_{r\downarrow}\rangle+\langle c_{r\uparrow}^{\dag}c_{r\uparrow}-c_{r\downarrow}^{\dag}c_{r\downarrow}\rangle$. Thus, if there exists a unit local moment, its magnetization should be one. Next, the double $f$-occupation number is $d_{f}(r)=\langle f_{r\uparrow}^{\dag}f_{r\uparrow}f_{r\downarrow}^{\dag}f_{r\downarrow}\rangle$, which should be vanished when interaction $U=\infty$ and reaches $1/4$ (recall half-filling condition) for non-interacting case. The last one is
$V_{cf}(r)=-\frac{1}{2}\sum_{\sigma}\langle f_{r\sigma}^{\dag}(c_{r+1,\sigma}-c_{r-1,\sigma})\rangle$,
denoting the strength of hybridization between conduction and local electron. Usually, one may use this to measure the strength of Kondo screening. ($c$-$f$ spin correlation function like $\langle\vec{S}^{f}(i)\cdot\vec{S}^{c}(j)\rangle$ can also be used to detect the strength of Kondo screening but our choice of $V_{cf}$ are physically more intuitive with the conventional wisdom.\cite{Paiva2003})

Now, from Figure~\ref{fig:compare}, results at $T=1/30$ and $T=0$ are basically indistinguishable, which verifies the finite temperature simulation presented in this manuscript.

\section*{Results}\label{sec_3}
\subsection*{Double occupation number}
\begin{figure}[!tb]
  \centering
  \includegraphics[width=0.4\columnwidth]{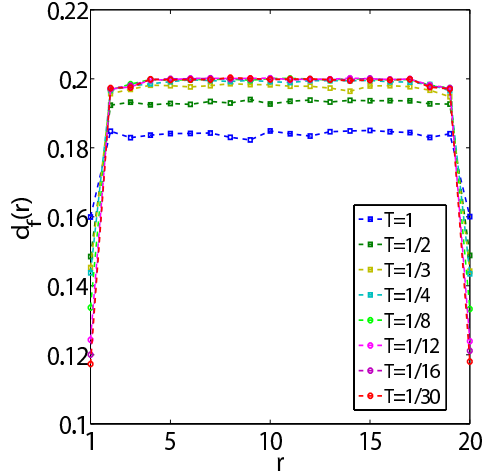}
\caption{Double occupation number of $f$-electron $d_{f}$ at different temperatures. When $T<1/4$, $d_{f}$ of the bulk sites ($5$-$15$th sites) has already reached its low-$T$ value. This implies that the charge fluctuation in bulk sites is saturated at lower temperature and in this strong charge fluctuating background ($d_{f}\sim0.2$ near its non-interacting value), the effect of spin fluctuation is not manifested at low temperature regime.
In contrast, $d_{f}$ of edge or boundary sites do not saturate
well below $T=1/4$ and it decreases down to lowest temperature. This means the charge fluctuation is much weaker than its bulk counterpart and the spin fluctuation should dominate the low temperature physics.}
\label{fig:double}
\end{figure}
The first quantity we examine is the double occupation number $d_{f}$ for local $f$-electron. Generally, we can read of the local charge fluctuation from $d_{f}(r)$. If charge degree of freedom can fluctuation freely like non-interacting electron system, $d_{f}(r)$ should approach $1/4$ for a half-filled system
while repulsive interaction reduces it and leads to a vanishing value for infinite interaction limit.
In addition, for a finite temperature system, thermal excitation should leads to a non-zero $d_{f}(r)$ even for very strong interaction case.

Here, from Figure~\ref{fig:double}, we observe several interesting features: 1) When $T<1/4$, $d_{f}$ of the bulk sites ($5$-$15$th sites) has already reached its low-$T$ value. This means that the charge fluctuation in bulk sites is saturated at lower temperature and in this strong charge fluctuating background ($d_{f}\sim0.2$ near its non-interacting value), the effect of spin fluctuation is not manifested at low temperature regime. 2) In contrast, $d_{f}$ of edge or boundary sites do not saturate
well below $T=1/4$ and it decreases down to lowest temperature. This tells us that the charge fluctuation is much weaker than its bulk counterpart and the spin fluctuation should dominate the low temperature physics. In other words, the edge is more susceptible than the bulk and the interaction effect would be more visible if we examine quantity related to the edge sites.

\subsection*{Site-resolved magnetization}

\begin{figure}[!tb]
\centering
  \includegraphics[width=0.4\columnwidth]{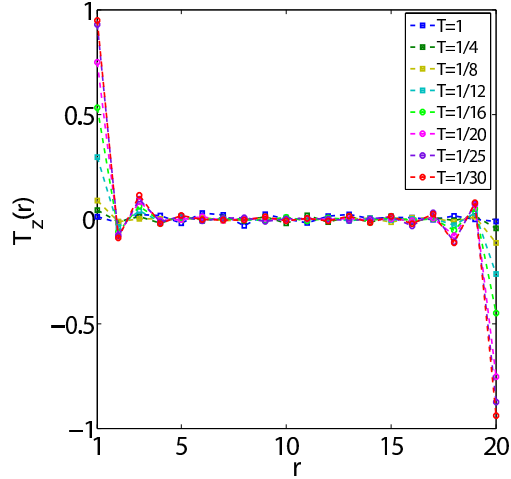}
\caption{Site-resolved magnetization $T_{z}$ versus site $r$ at different temperatures. When temperature is high ($T>1/4$), magnetization on each site is small and no local moment appears at edge sites or bulk sites. However, if further decreasing temperature, noticeable magnetization gradually develops at edge ($T\leq1/8$) while the bulk is still non-magnetic. Furthermore, at lowest temperature, the edge magnetization seems to saturate to unit local moment, indicating the development of Haldane phase.}
\label{fig:2}
\end{figure}

Next, in Figure~\ref{fig:2}, we have shown the site-resolved magnetization $T_{z}$ versus site $r$ at different temperatures from high temperature $T=1$ down to low-temperature regime $T=1/30$.

From Figure~\ref{fig:2}, it is clear that when temperature is high ($T>1/4$), magnetization on each site is small and no local moment appears at edge sites or bulk sites. However, if further decreasing temperature, noticeable magnetization gradually develops at edge ($T\leq1/8$) while the bulk is still non-magnetic. Furthermore, at lowest temperature, the edge magnetization seems to saturate to unit local moment, which is identical to the free $S=1/2$ edge spin found in our previous zero-temperature quantum Monte Carlo simulation and also confirmed by an independent density-matrix-renormalization-group calculation.\cite{Zhong2017,Lisandrini2017}

Because the non-vanishing edge magnetization is believed to be a fundamental signature of Haldane phase in both pure spin systems and charge doped models,\cite{Wen2016,Nourse2016,Lobos2015,Mezio2015,Hagymasi2016}
the temperature dependence of edge magnetization implies that the topological Haldane phase is stable at finite temperature. Physically, if the bulk gap does not close/filled by thermal effect, the topological feature of Haldane phase, e.g. free local moment at edge, should not be changed. Although we cannot give quantitative value of the bulk gap, it could be estimated as the Kondo energy scale
$T_{K}\sim te^{-Ut/V^{2}}\sim\mathcal{O}(1/10)$. Therefore, we understand that edge magnetization will establish when $T<T_{K}$ is reached and it is indeed the case shown in Figure~\ref{fig:2}.

\subsection*{Spin structure factor}
Alternative, one may use other physical observable to detect the free local moment in Haldane phase.
Here, the temperature dependent spin structure factor $S_{\mathrm{spin}}$ will be inspected. Specifically, it is define by
\begin{eqnarray}
S_{\mathrm{spin}}=\frac{1}{L}\left\langle\left(\sum_{r}2S_{z}^{c}(r)+\sum_{r}2S_{z}^{f}(r)\right)^{2}\right\rangle
=\frac{4}{L}\sum_{r,l}[\langle S_{z}^{c}(r)S_{z}^{c}(r)\rangle+2\langle S_{z}^{c}(r)S_{z}^{f}(r)\rangle+\langle S_{z}^{f}(r)S_{z}^{f}(r)\rangle]
=S_{\mathrm{spin}}^{cc}+2S_{\mathrm{spin}}^{cf}+S_{\mathrm{spin}}^{ff},
\end{eqnarray}
where $S_{\mathrm{spin}}^{cc},S_{\mathrm{spin}}^{ff},S_{\mathrm{spin}}^{cf}$ denote contribution from
conduction electron, local $f$-electron and their interplay, respectively. More importantly, by definition, one can read out total local moment $M_{z}$ from $S_{\mathrm{spin}}$ by just multiplying the size of system, namely
$\sum_{r}(m_{z}(r))^{2}=M_{z}^{2}=L\times S_{\mathrm{spin}}$.\cite{Vekic1994}
Here, we have intuitively introduced the site-dependent local moment $m_{z}(r)$, which will be useful to explain emergent edge local moments.

\begin{figure}[!tb]
\centering
  \includegraphics[width=0.6\columnwidth]{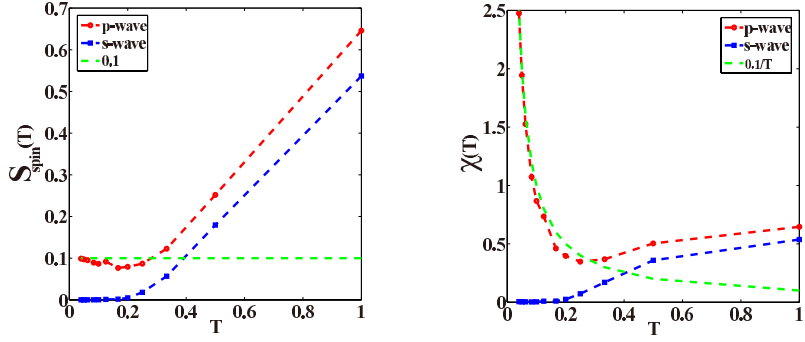}
\caption{(Left) Spin structure factor $S_{\mathrm{spin}}$ versus temperature $T$. The green dash line is the guide to value $0.1$, which is approached at low temperature. (Right) Static spin susceptibility $\chi$ versus temperature $T$, which shows a $\frac{0.1}{T}$ behavior at the low-$T$ limit.}
\label{fig:4}
\end{figure}

Now, turn our attention to Figure~\ref{fig:4}, we see that the spin structure factor $S_{\mathrm{spin}}$ approaches $0.1$ at low temperature ($T\leq1/12$) for our $p$-wave periodic Anderson model $H_{p-wave}$ (Equation~\ref{eq1}). Recall that our system has size $L=20$, so we find
$M_{z}^{2}=2$ and this exactly corresponds to the summation over two free unit local moments. Obviously, these two moments should be the expected edge magnetization of local moment. Therefore, the temperature dependent spin structure factor also gives the information of forming of edge magnetization and this is consistent with findings in last subsection. For comparison, the spin structure factor for $s$-wave model $H_{s-wave}$ (Equation~\ref{eq2}) is also shown and it decays to zero as expected since the $s$-wave one does not support free edge local moments and all spin moments are quenched by collective Kondo screening effect.

Correspondingly, the spin susceptibility is related to the spin structure factor as $\chi(T)=\frac{S_{\mathrm{spin}}}{T}$.
For free magnetic moments, it gives rise to a Curie-like behavior at low temperature, which is exactly seen in the right plot in Fig.~\ref{fig:4}, where our $p$-wave model has the expected $1/T$-law while the $s$-wave counterpart vanishes at low temperature due to quenched local moments.

\subsection*{$c$-$f$ hybridization and its implication}
Finally, we examine the behavior of $c$-$f$ hybridization $V_{cf}$, which shows the local information of Kondo screening.

\begin{figure}[!tb]
\centering
  \includegraphics[width=0.5\columnwidth]{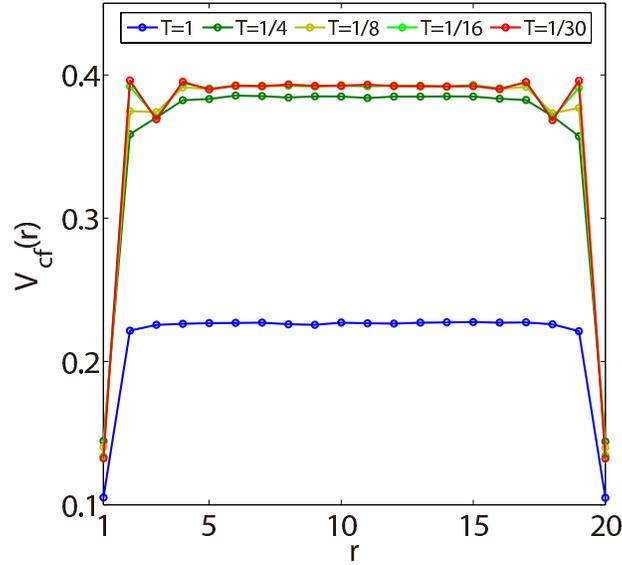}
\caption{Site-resolved $c$-$f$ hybridization $V_{cf}$ versus site $r$ at different temperatures $T$. Note that $V_{cf}$ is heavily reduced but still not vanishing at edge sites.}
\label{fig:3}
\end{figure}

From Figure~\ref{fig:3}, it is found that at low temperature the edge $c$-$f$ hybridization is much smaller than its bulk counterpart ($0.1$ versus $0.4$), which states the Kondo screening is heavily weakened at edge sites and instead, the spin degree of freedom may have chance to form local moments. Notice that at $T=1/4$, its edge $c$-$f$ hybridization has in fact near the value of $T=1/30$ case. At the same time, edge magnetization at $T=1/4$ is nearly vanishing from Figure~\ref{fig:2}, thus it seems to give us a flavor of possible temperature-driven Kondo breakdown at intermediate temperature regime .\cite{Alexandrov2015}

However, such attractive phenomena of Kondo breakdown are hard to detect in numerical simulations because there is no suitable 'order parameter' for describing the Kondo screening (let alone the Kondo breakdown) and it is well-known that for the impurity problem, the high temperature local moment regime gradually crossovers into the low temperature Kondo screening regime without observable singularity.\cite{Coleman2015} The Kondo screening in lattice problem seems to be more complicated and some recent angle-resolved photoemission spectroscopy (ARPES) experiments even challenge our basic picture on Kondo lattice systems.\cite{Kummer2015,Chen2017}(e.g. Is the renormalized quasi-particle band really responsible for heavy fermion liquid? Is there a true small-to-large Fermi surface reconstruction in Kondo lattice compound when decreasing temperature?)

\section*{Discussion}
\label{sec_4}

\subsection*{An emergent energy scale: the crossover temperature $T_{cr}$}
From Figure~\ref{fig:4}, we have seen that the spin structure factor and susceptibility begin to show the  developing (for $p$-wave hybridization) or quenching (for $s$-wave hybridization) of free local moments at $T=1/4$. Meanwhile, from Figure.~\ref{fig:double} and \ref{fig:3}, below $T=1/4$, both the double occupation number and $c$-$f$ hybridization of the bulk sites are near their low temperature value. In addition, when $T<1/4$, the edge magnetization starts to develop as what can be seen in Figure~\ref{fig:2}.

Given these observations, we are able to define a crossover temperature $T_{cr}$, which marks the crossover from non-universal high temperature regime to interesting low temperature regime accompanied   by sharp changes of many observables. But, what is $T_{cr}$? We may guess it by listing possible candidates like Kondo scale $T_{K}\sim te^{-Ut/V^{2}}\sim\mathcal{O}(1/10)$, Kondo coupling $J_{K}\sim t^{2}/U\sim1/2$ and RKKY coupling $T_{RKKY}\sim J_{K}^{2}/t\sim1/4$. Thus, it seems that the crossover temperature $T_{cr}$ can be ascribed to be the RKKY energy scale $T_{RKKY}$, which is well consistent with phenomenological two-fluid theory proposed to re-understanding the heavy fermion systems.\cite{Yang2016}

Additionally, when $T<T_{cr}$, the $c$-$f$ hybridization (e.g. Figure~\ref{fig:3}) shows an extra oscillation feature around boundary, which may reflect the emergence of the underlying oscillating RKKY interaction.

\subsection*{Comparison with conventional $s$-wave periodic Anderson model}
\begin{figure}[!tb]
\centering
  \includegraphics[width=0.6\columnwidth]{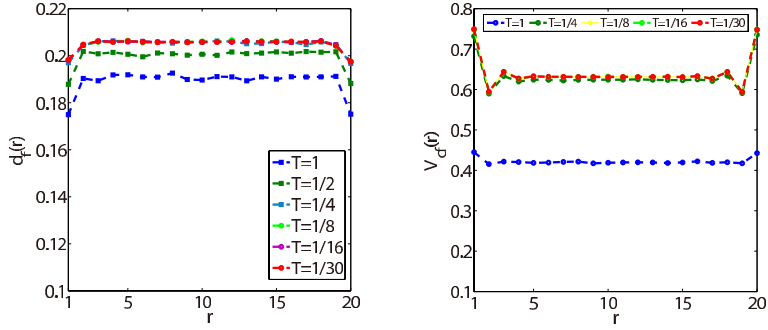}
\caption{(Left) Double occupation number of $f$-electron $d_{f}(r)$ and (right) $c$-$f$ hybridization $V_{cf}(r)$ for conventional $s$-wave periodic Anderson model. Below $T_{cr}=1/4$, both $d_{f}(r)$ and $V_{cf}(r)$ have nearly approached their low-$T$ limit value. The double occupation number at boundary is larger than the $p$-wave case, which is due to the more connectivity of $s$-wave (on-site) hybridization. In contrast to $p$-wave model, the edge $c$-$f$ hybridization of $s$-wave model is enhanced compared with its bulk sites, which results from the same reason of double occupation number.}
\label{fig:41}
\end{figure}

In this subsection, since the spin structure factor and spin susceptibility of $s$-wave model $H_{s-wave}$ (Equation~\ref{eq2}) have been analyzed in last subsection, we here present temperature-dependent double $f$-occupation number and $c$-$f$ hybridization. It should be emphasized that the site-resolved magnetization are all zero and we will not show it here, which is quite different from the $p$-wave model $H_{p-wave}$ (Equation~\ref{eq1}) discussed lengthly in the main text, where finite edge magnetization has been found at low temperature. In Figure~\ref{fig:41}, we first note that below crossover temperature $T_{cr}=1/4$, observables have nearly approached their low-$T$ limit. Then, it is seen that the double occupation number at boundary is larger than the $p$-wave case. This should be due to the more connectivity of $s$-wave (on-site) hybridization. Moreover, in contrast to $p$-wave model, the edge $c$-$f$ hybridization of $s$-wave model is enhanced compared with its bulk sites, which simply results from the same reason of double occupation number.

\section*{Conclusion}
\label{sec_5}

In conclusion, we have studied the $p$-wave periodic Anderson model at finite temperature in terms of numerically exact DQMC simulation.
It is found that with the non-local $p$-wave hybridization, the Haldane phase is a stable state of matter at low temperature as seen from temperature dependent quantities like edge magnetization, spin structure factor and static spin susceptibility. A crossover temperature related to RKKY interaction seems to be identified, which gives a good example for the promising heavy fermion two-fluid theory.
As for the temperature-driven surface Kondo breakdown, it is immature to identify its existence or not due to its intricate nature and beyond the scope of our present work.
Therefore, more works on surface Kondo breakdown or more general the (bulk) Kondo breakdown mechanism
are desirable,\cite{Si2001,Senthil2004,Paul2007,Zhong2012} which would open new realm in the heavy fermion community and other strongly correlated electron systems.

Because our model here is a one-dimensional model, we cannot give direct implications to realistic three-dimensional topological Kondo insulator materials like SmB$_{6}$. Nevertheless, from our extensive numerical simulations, we learn that the electron correlation effect is more important for edge of system than the bulk. To our surprise, very recent measurements of SmB$_{6}$ on specific heat, magnetic
quantum oscillations and thermal conductivity all indicate a bulk anomaly, i.e. there exists a bulk Fermi
surface in this Kondo insulating compound.\cite{Hartstein2017} It seems that electron correlation effect is also crucial for bulk properties of realistic high-dimensional topological Kondo insulator Therefore, an unbiased DQMC simulations of two or three-dimensional p-wave periodic Anderson model are highly desirable.
However, due to intrinsic spin-dependent $p$-wave hybridization in those models,\cite{Coleman2015} severe fermion minus-sign problem will occur, which prohibits the reliable simulations of DQMC even at half-filling condition.

Finally, since ultracold atom setups have been good laboratory for condensed matter physics and several proposals have been made for simulating ubiquitous heavy fermion phenomena,\cite{Nakagawa2015,Zhong2017F,Caro2017} we expect the temperature effect studied in the present work may be observed in forthcoming state-of-art experiments.

\section*{Acknowledgments}
This research was supported in part by NSFC under Grant No.~$11325417$, No.~$11674139$, and No.~$11704166$, the Fundamental Research Funds for the Central Universities, Science Challenge Project under Grant No. JCKY2016212A502, SPC-Lab Research Fund (NO. XKFZ201605)
and the Foundation of LCP.

\section*{Author contributions statement}

Y.Z. designed the work and carried out the calculations. Y.Z., Y.L., Q.W., K.L. and H.L. analysed the data and wrote the manuscript. Y.Z. wrote the DQMC code. All authors revised the manuscript.

\section*{Additional information}
\subsection*{Competing financial interests:} The authors declare no competing financial interests.
\subsection*{Data availability statement:}
The datasets generated during and/or analysed during the current study are available from the corresponding author on reasonable request.
\end{document}